\documentclass[a4paper,11pt]{article}
\usepackage{pos}

\usepackage[numbers]{natbib}
\usepackage{cleveref}
\hypersetup{colorlinks=true,linkcolor=blue,citecolor=magenta,filecolor=magenta, urlcolor=cyan}

\def\ie{{\it i.e.}}
\def\eg{{\it e.g.}}
\newcommand{\ch}{{\sc CalcHep}}
\newcommand{\fr}{{\sc FeynRules}}
\newcommand{\ma}{{\sc MadAnalysis}~5}
\newcommand{\maddm}{{\sc MadDM}}
\newcommand{\mg}{{\sc MG5\_aMC}}
\newcommand{\micromegas}{{\sc Mi\-crO\-ME\-GAs}}
\newcommand{\darksusy}{{\sc DarkSUSY}}

\newcommand{\py}{{\sc Pythia}~8}

\newcommand{\superiso}{{\sc SuperIso Relic}}

\newcommand{\sigmav}{{\langle \sigma v\rangle}}

\title{Review on Dark Matter Tools}

\author*[a]{Chiara Arina}

\affiliation[a]{Centre for Cosmology, Particle Physics and Phenomenology (CP3), IRMP, UCLouvain,\\
  Chemin du Cyclotron 2, 1348 Louvain-la-Neuve, Belgium}

\emailAdd{chiara.arina@uclouvain.be}

\abstract{
Whilst the need for dark matter was established almost a century ago, only its gravitational interaction has been confirmed so far, allowing for plethora of models for dark matter. The Weakly Interacting Massive Particles (WIMPs) category has received by far the biggest attention, however despite the enormous experimental efforts, these particles remain elusive. The attention of the community has hence moved on to investigate the dark matter landscape over a much larger number of models with varying degrees of resemblances and differences in their predictions. This calls for the need to organise the various facets of dark matter models and their signatures, in order to maximise the experimental sensitivity and to select the models which are compatible with existing data. In this paper, I provide a short review of the most widespread public codes capable of computing dark matter observables. In particular, I discuss what is the status of each numerical tool in terms of: (i) capturing the WIMP phenomenology 
and (ii) accounting for new trend dark sector models that might be weakly coupled to ordinary matter and/or be strongly self-interacting. This short review has the aim of guiding the user towards selecting the best suited public code to confront his/her model with the largest variety of theoretical predictions and experimental data in order to determine the parameter space consistent with observations for his/her favourite dark matter model.
}

\FullConference{%
  Tools for High Energy Physics and Cosmology - TOOLS2020\\
  2-6 November, 2020\\
  Institut de Physique des 2 Infinis (IP2I), Lyon, France}

\begin{document}
\maketitle

\section{Introduction}

The quest for dark matter has started almost a century ago, however at present very little is known about its nature. Its presence is strongly required for instance by the cosmological model $\Lambda \rm CDM$~\cite{Aghanim:2018eyx}, structure formation~\cite{Navarro:2008kc,BoylanKolchin:2009nc,Vogelsberger:2014dza}, galaxy rotation curves, see \ie~\cite{Bovy:2012tw,Bertone:2010zza}. The astrophysical and cosmological hints can however only probe the gravitational interaction of dark matter and provide no insights in other types of interactions, in particular with the Standard Model (SM) particles. 

Under the assumption that dark matter particles do interact with SM particles, the most studied candidates so far are the so-called Weakly Interacting Massive Particles (WIMPs). 
These particles feature interaction strength of the order of the weak force and consequently achieve naturally the correct relic density in the GeV and TeV mass range. Besides the fact that they arise in many models beyond the SM, such as supersymmetry (SUSY) or extra-dimension, see \eg~\cite{Bertone:2010zza}, they result in a variety of signals at accessible energy scales. A multitude of experimental approaches has been undertaken to detect WIMPs, ranging from dark matter searches using underground detectors (direct detection), to observations of gamma rays, cosmic rays and neutrinos in astrophysical environments (indirect detection), and searches for missing energy signals at colliders (production). Yet, despite the enormous experimental effort, WIMPs remain elusive. Even though they are not excluded, see \eg~\cite{Leane:2018kjk}, new directions are explored, such as dark hidden sectors, Feebly Interacting Massive Particles (FIMPs)~\cite{Bernal:2017kxu}, sub-GeV dark matter candidates~\cite{Battaglieri:2017aum}. 
For instance, these latter imply that dark matter direct detection arises via electronic recoils (instead of the standard nuclear recoils), while the former may imply very weakly coupled particles to the SM but strongly self-interacting and decoupling from the thermal bath with a temperature different than to one of the SM particles. The FIMPs encompass dark matter candidates in all the mass range between sub-GeV to TeV. These particles are feebly coupled to the SM and for this reason not in thermal equilibrium in the early universe, but produced by the freeze-in mechanism instead, see~\cite{McDonald:2001vt,Hall:2009bx} for details. These are only few examples to make the point that these new avenues require  new theoretical frameworks for the predictions and consequently call for new numerical routines to compare those predictions with the experimental results in the spirit of global fits. 

From the software point of view, this last decade is characterised by constant efforts to improve the public codes for dark matter observables along three main directions: (i) cover all WIMP phenomenology and improve its accuracy with the inclusion of next-to-leading order (NLO) corrections; (ii) include the new phenomenology arising in the new dark matter paradigm and (iii) make the public code portable/well linked with other available software. This should help the user to perform global predictions and sampling of the parameter space, simultaneously taking into account constraints from collider and different dark matter probes. Even though a bit aside from the scope of this review, point (iii) is of the uttermost importance and few comments are in order, since it is developed in several directions. On the one hand,  \fr~\cite{Alloul:2013bka}, \textsc{SARAH}~\cite{Staub:2012pb} and \textsc{LanHEP}~\cite{Semenov:2014rea} allow the user to implement his/her dark matter model at the level of the Lagrangian, and produce output files for several high energy public tools, including \micromegas\, and \maddm\, for computing dark matter observables. Several dark matter models are already publicly available in the \fr\, database\footnote{\tiny{\url{https://feynrules.irmp.ucl.ac.be/wiki/ModelDatabaseMainPage}}}, such as the so-called $s$- and $t$-channel simplified models\footnote{\tiny{\url{https://feynrules.irmp.ucl.ac.be/wiki/DMsimp} and \url{https://feynrules.irmp.ucl.ac.be/wiki/DMsimpt} for $s$- and $t$-channel respectively.}} used at the LHC, see~\cite{Abercrombie:2015wmb,Backovic:2015soa,Boveia:2016mrp,Arina:2018zcq,Arina:2020udz}. From the same model Lagrangian, \fr\,  provides output files for computing collider predictions with \mg~\cite{Alwall:2011uj} (UFO format~\cite{Degrande:2011ua}, the same used in \maddm) and to apply collider bounds with the recasting tool \ma~\cite{Conte:2012fm}. Alternatively, from \micromegas \, it is straightforward to compute collider exclusion bounds in the simplified model framework by using the SModelS tool~\cite{Ambrogi:2017neo}. These chains of tools are pretty well established and have led to numerous publications, see \eg~\cite{Arina:2016cqj,Barducci:2016pcb,Arina:2020tuw,Goodsell:2020lpx}. On the other hand, the GAMBIT collaboration~\cite{Athron:2017ard} provides various modules for dark matter observables~\cite{Workgroup:2017lvb}, cosmology~\cite{Renk:2020hbs}, sampling~\cite{Workgroup:2017htr}, to perform consistently global analysis of dark matter models, see \eg~\cite{Athron:2017qdc,Kvellestad:2019vxm,Athron:2020maw,Arina:2019tib}.

In this review I will focus specifically on dark matter tools and consider only the four major public and generic codes: \superiso\footnote{\tiny{\url{http://superiso.in2p3.fr/relic/}}}~\cite{Arbey:2018msw}, \darksusy\footnote{\tiny{\url{https://darksusy.hepforge.org/}}}~\cite{Bringmann:2018lay}, \micromegas\footnote{\tiny{\url{https://lapth.cnrs.fr/micromegas/}}}~\cite{Belanger:2018ccd} and \maddm\footnote{\tiny{\url{https://launchpad.net/maddm}}}~\cite{Ambrogi:2018jqj,Arina:2020kko}. There are many more specific dark matter tools available on the market, which I will mention whenever relevant. In~\cref{sec:sofar}, I will briefly review the current status of these numerical tools in relation with the coverage of the WIMP phenomenology. Subsequently in~\cref{sec:features}, I will discuss the most salient features that go beyond the WIMP scenario for each tool separately.  My concluding remarks are discussed in~\cref{sec:concl}. This short paper is not meant to give an exhaustive description of the new avenues towards dark sectors nor an in-depth description of the public codes, for which I refer to the documentation provided by the authors of the software.
This review is rather meant to highlight features that distinguish the different numerical tools, in such a way to guide the user towards the most suitable code for his/her purpose.

\section{WIMPs: where do the tools stand so far?}\label{sec:sofar}

The development of public generic tools has started already at the beginning of the 21st century: \micromegas~\cite{Belanger:2001fz} is the first code being developed, followed by \darksusy~\cite{Gondolo:2004sc}, \superiso~\cite{Arbey:2009gu} and then subsequently by \maddm, which is by far the more recent one~\cite{Backovic:2013dpa}. As all codes are on the market from a few years now, they all cover the necessary basics of the WIMP phenomenology, to various extent.

\paragraph{Relic density} WIMPs achieve the correct relic density via the freeze-out mechanism. The dark matter particles have interactions of the order of the electroweak force, which keep them in chemical (and kinetic) equilibrium in the thermal bath enough to decouple when they are non-relativistic and at the photon temperature for which the thermally averaged cross-section $\sigmav$ is around $3 \times 10^{-26} \rm cm^3/s$ (for the $s$-wave annihilation at low relative velocity $v$). This ensures that the relic abundance is close to $\Omega h^2 = 0.12$, value measured by Planck~\cite{Aghanim:2018eyx}. Departing from this simple case, dark matter particles can feature a $\sigmav$ which varies rapidly with energy close to resonances or to thresholds of new annihilation channels, or the model can have coannihilation. The latter arises when there are new particles beyond the SM ones, which are odd under the dark group similarly to the dark matter particle, and which are close in mass to the dark matter particles (namely 10\% - 15\% in mass splitting at most). These heavier particles are present in the thermal bath at the time of dark matter freeze-out and their annihilations contribute to it. To account then for all annihilation and coannihilation processes it is customary to define an effective thermally averaged cross-section $\sigmav_{\rm eff}$, see \ie~\cite{Gondolo:1990dk,Edsjo:1997bg}. 

All public codes provide the exact relativistic single-integral formula for $\sigmav_{\rm eff}$ and solve the corresponding Boltzmann equation to find the dark matter yield at present time. More specifically, \superiso\, has included analytically all the hundreds of diagrams relevant for neutralino annihilation and coannihilation in supersymmetric scenarios such as the MSSM. This is similar to the case of \darksusy\, which was born for supersymmetry, even though the user can directly provide $\sigmav$ for his/her model in order to compute $\Omega h^2$, which is in addition determined by using the most precise treatment described in~\cite{Steigman:2012nb}. \micromegas\, and \maddm\, rely on the on-the-fly computation of all matrix elements by \ch~\cite{Belyaev:2012qa} and \mg\, respectively, including coannihilation processes; they additionally handle the case of multi-component dark matter. Furthermore, the user can improve $\sigmav$, such as with NLO terms, by editing a simple predefined function within \micromegas. There are indeed works that have shown that NLO corrections are relevant, for instance when the annihilation cross-section is suppressed by $d$-wave ($v^4$) or in supersymmetric models, see \eg~\cite{Giacchino:2015hvk,Beneke:2016ync}. In the case of supersymmetry, few specific packages have been developed, such as DM@NLO~\cite{Herrmann:2007ku}, which are typically compatible with \micromegas. \darksusy\, allows for the computation of the kinetic decoupling of dark matter particles, which happens much after they have frozen-out (chemical decoupling). This is phenomenologically relevant as it firstly includes higher moments in the Boltzmann equation, which can considerably change the relic density prediction in some corners of the model parameter space, and secondly sets a relevant scale for the formation of small-scale structures with observational consequences in indirect detection signals, see~\cite{Bringmann:2006mu,Binder:2017rgn}.

\paragraph{Direct detection}
Standard dark matter direct searches measure the recoil energy of nuclei hit by dark matter particles crossing (large) detectors deep underground. The typical recoil energy is at the keV energy scale, as the dark matter particles are non-relativistic, have a velocity of about 220 km/s which lead to a momentum transfer in the scattering of the order of MeV, see \eg~\cite{Gould:1988eq,Cerdeno:2010jj,Arina:2013jma,Schumann:2019eaa} for the generalities. Because of this small velocity, typically all terms proportional to the momentum transfer and to the velocity, are neglected and only the leading spin-independent and spin-dependent interaction terms are kept. This is the approach followed by \micromegas\, and \maddm\, to compute the elastic cross-section for WIMP-nuclei scattering at both leading order and NLO in the case of the former code. The NLO corrections in general take into account the contribution of gluon operators (called twist-2 operators), see for details~\cite{Hisano:2015bma}.

As shown by an effective non-relativistic field theory approach, the actual basis of operators that describe dark matter particles interacting with nuclei contains around 15 terms, and most of them do depend on the velocity or the momentum transfer~\cite{Fan:2010gt,Cirigliano:2012pq,Fitzpatrick:2012ix}. When the dominant spin-independent and/or spin-dependent operators are zero, subdominant terms become relevant and might be in the sensitivity range of current experiments, see \eg~\cite{Arina:2014yna,Bozorgnia:2018jep}. \darksusy\, has recently implemented the velocity dependence for the elastic scattering of dark matter. 

\superiso, \micromegas~\cite{Belanger:2020gnr} and \maddm\,  have recasted few experimental likelihoods or upper bounds to allow the user to directly compare the theoretical predictions with the most stringent experimental limits, such as  XENON1T~\cite{Aprile:2018dbl} for spin-independent and PICO-60~\cite{Amole:2017dex} for spin-dependent scattering. \superiso, \micromegas\, and \darksusy\,  allow as well for modifications of the dark matter velocity distribution, to account for astrophysical uncertainties. Nuclear uncertainties are taken into account in all the tools by allowing the user to easily change the value of the nuclear form factors by hand.

There are also public codes devoted only to direct detection searches, such as {\sc{DDCalc}}, included in {\sc{DarkBit}}~\cite{Workgroup:2017lvb}, {\sc{RAPIDD}}~\cite{Cerdeno:2018bty}, which is based on the non-relativistic effective field theory description of dark matter interactions with nuclei, and {\sc RunDM}~\cite{DEramo:2016gos}. The latter takes into account the running of the couplings from the UV scale of the model down to the keV-MeV scale of direct detection experiments.  

\paragraph{Indirect detection}
Dark matter indirect searches look for WIMP annihilation (or decay) products in dense astrophysical environments, see \eg~\cite{Cirelli:2012tf,Gaskins:2016cha} for general reviews on the matter. In particularly, they look for an excess (with respect to the astrophysical background) in the flux of gamma rays, cosmic rays (positrons and anti-deuterium mainly) and neutrinos from the galactic centre, from dwarf spheroidal galaxies, from the Sun, namely everywhere the dark matter density is large and/or dominates over the baryonic component. 

From the particle physics point of view, there are two key quantities relevant to obtain predictions for the expected flux: the velocity averaged annihilation cross-section $\sigmav_{\rm ann}$ and the energy spectrum $dN/dE$ of the final products listed above. 
The computation of the matrix element leading to $\sigmav_{\rm ann}$ proceeds in the same way as for the relic density in each tool, while the energy spectra are tabulated in all codes for two body final states. \maddm provides the tabulated energy spectra, imported from \textsc{PPPC4DMID}~\cite{Cirelli:2010xx}, which effectively speeds up the computation of the expected flux, when the dark matter annihilates into the usual SM final states. Additionally \maddm\, allows the user to compute the energy spectra for any final state with $n$ particles, independently of their nature (SM or beyond), by automatically running \py~\cite{Sjostrand:2014zea}. The energy spectra can change significantly if electroweak corrections are included, for which two treatments at the time of writing are available: \textsc{PPPC4DMID}, that is based on~\cite{Ciafaloni:2010ti}, and ref.~\cite{Bauer:2020jay}. Additionally, the shape of the energy spectrum can acquire spectral features if  internal~\cite{Bringmann:2012ez} or electroweak~\cite{Bringmann:2013oja} Bremsstrahlung is considered ($2 \to 3$ processes). The first is automatically included in \micromegas\, while both are encoded in \darksusy\,  for specific models. \maddm\, can compute any of those corrections for a generic model if this is explicitly required by the user.

From the astrophysical point of view, indirect detection requires a lot of inputs and development. First, the energy spectrum is produced at the source; in case of charged particles, which diffuse in the galaxy, a propagation code is needed to obtain the correct spectrum at detection. Cosmic-ray propagation is available in all codes but \maddm, where instead an interface to the numerical code \textsc{DRAGON}~\cite{Evoli:2008dv} is provided. Gamma rays and neutrinos do not suffer propagation effects and point straight to the source where they were produced, hence their energy spectra is the same at source and detection point (modulo neutrino oscillations). Concerning the latter, one of the most promising searches is to look for neutrinos from the Sun and/or from the Earth~\cite{Gould:1987ir} with IceCube~\cite{Aartsen:2016zhm}: these predictions are fully available in \micromegas\,  and \darksusy\, with a high level of precision. These predictions are rather involved as their require a modelling of the Sun and Earth together with the computation of neutrino oscillations in matter. 
Concerning gamma rays, one of the most sensitive search is the Fermi-LAT search from dwarf spheroidal galaxies~\cite{Fermi-LAT:2016uux}. For this search, \maddm\, and \superiso\, not only compute the expected flux but provide as well a recasted Fermi-LAT likelihood analysis to compare directly the model parameter point with the experimental data. \superiso\, in this respect is the public code that has the most advanced experimental module, as it provides as well a recasting of the AMS-02 antiproton analysis~\cite{PhysRevLett.117.091103}.

There are many public codes devoted to the study of astrophysical signals from dark matter, which are typically model-independent. Namely, they focus on the precise description of the astrophysical flux produced by dark matter and by the astrophysical background and they require the user input concerning the particle physics model. A non-exhaustive list is given by:  \textsc{gamLike} included in {\sc{DarkBit}}, \textsc{GALPROP}~\cite{Vladimirov:2010aq}, \textsc{DRAGON}, {\sc{PPPC4DMID}}, {\sc{MADHAT}}~\cite{Boddy:2019kuw}, {\sc{CLUMPY}}~\cite{Bonnivard:2015pia}, $\chi$aro$\nu$~\cite{Liu:2020ckq} and {\sc{USINE}}~\cite{Maurin:2018rmm}.

\vspace{0.5cm}
The output for relic density, direct detection and indirect detection of the different public codes has been compared in various configurations and with very different dark matter models. Overall, the results for the theoretical predictions generally agree very well, the discrepancies being never more than 10\%, which turns out to be of the order of the theoretical error. Sometimes, larger discrepancies are found, but usually, when this happens, the causes are well understood and trackable. For instance, in refs.~\cite{Ambrogi:2018jqj,Arina:2020udz} the discrepancies in the relic density calculation arise in the different treatment of the QCD sector in \mg\,   and \ch. In summary, in the standard WIMP freeze-out scenario, all public codes have been deeply tested and compared, process that has allowed to solve numerous bugs.

\section{Tell me your model, I will tell you the tool}\label{sec:features}

The widening of the dark matter landscape has broaden the spectrum of theoretical predictions for relic density, direct and indirect detection and to a minor extent of LHC dark matter searches. In order to efficiently describe the phenomenology of the new avenues, the public tools are constantly evolving to encompass as much as possible the new predictions. I list below the remarkable features of each of the four numerical tools described in this review. While for WIMPs the various codes are more or less equivalent, it appears clear in the following that for the new avenues each one of them has definite characteristics  that render it unique to the user.

\subsection{\superiso: So far it's all about supersymmetry}
The current \superiso\, version v4 is a mixed Fortran/C code which allows to compute comprehensively observables in supersymmetry, together with an experimental module that enables the user to automatically confront his/her model with experimental data for direct and indirect detection, as already described above. It is actually one of the reference codes to perform  statistical comparisons between the SM and supersymmetric models in the light of recent collider, high energy, flavour physics and dark matter data, see \eg~\cite{Hryczuk:2019nql}.

Besides dealing with the supersymmetric WIMP phenomenology, \superiso\, can be easily interfaced with \textsc{AlterBBN}~\cite{Arbey:2018zfh}, a stand-alone public tool which allows to compute Big Bang Nucleosynthesis observables in standard and modified cosmologies. The combination of these two tools permits to test the supersymmetric dark matter hypothesis in alternative cosmologies, where for instance the presence of quintessence or scalar fields modifies the expansion rate of the universe and consequently the relic abundance of the neutralino, see~\cite{Arbey:2018uho}. 

Lastly, \superiso\, is foreseen to be able to compute dark matter observable for generic dark matter models in a near future. This augmented capacity is based on the public tool \textsc{MARTY}~\cite{Uhlrich:2020ltd}, a C++ framework that automates calculations from the Lagrangian to physical quantities such as matrix elements or cross-sections. This framework can fully simplify physical quantities in a large variety of models, making possible to quickly compare a dark matter model with all accessible experimental data.

\subsection{\darksusy: Towards dark sectors and new interactions}
The current \darksusy\, version (v6) is a Fortran code with a clear modular structure; in each module, such as the MSSM module, or the generic WIMP module, the user can select and extract the functions he/she needs and link them to his/her model by providing few inputs, such as the dark matter mass, $\sigmav$, etc. \darksusy\, has implemented several features that allow the user to describe the phenomenology of generic dark sectors. 

Concerning the relic density, besides the precision routines described above for the standard freeze-out case, it computes the freeze-out and the corresponding thermal $\sigmav$ for dark matter particles in hidden sectors, which can have a temperature different than the photon temperature, see~\cite{Bringmann:2020mgx}.

Dark matter in dark sectors can have large self-interactions, which can be obtained in two ways:  (i) from strongly coupled dark sectors, similarly to QCD, or (ii) from weakly coupled theories featuring a very light mediating particle.  The former case, where the large self-interactions do not depend on the dark matter velocity, is severely bounded from galaxy cluster observations, see \eg\, the well-known constraints from the Bullet cluster~\cite{Markevitch:2003at}.  In the latter case, self-interactions  do depend on velocity and become stronger at smaller dark matter velocities. Phenomenologically, it is possible to have large effects on small scales consistent with the astrophysical constraints on larger scales. This is the case implemented in \darksusy~\cite{Bringmann:2016din}, which includes the computation of the Sommerfeld enhancement (see \eg~\cite{Iengo:2009xf,Iengo:2009ni,Arina:2010wv} for its numerical solution), cutoff at small scales due to kinetic decoupling and the momentum transfer cross-section $\sigma_T$ not only in the Born limit but accounting for non-perturbative effects.

Concerning direct detection, \darksusy\, has started to cover the phenomenology of light dark matter, the so-called sub-GeV dark matter. It includes `reverse' direct detection~\cite{Cappiello:2018hsu}, where the dark matter is now the target and ordinary matter is the beam. More specifically, the number density of sub-GeV dark matter is very high, as its mass is small, hence the cosmic-ray propagation can be affected when these charged particles scatter elastically off dark matter particles in the galaxy. Building on top of this, one of the \darksusy\, authors has shown that current large volume direct detection and neutrino experiments are sensitive to a new population of dark matter particles, which is boosted at high velocity by up scattering off cosmic-rays close to the Sun position~\cite{Bringmann:2018cvk,Plestid:2020kdm}. The related theoretical predictions are available within the code.
Lastly, notice that the interest in sub-GeV dark matter is spreading in the scientific community and other efforts are ongoing. For instance, a specific tool, Damascus~\cite{Emken:2017qmp}, that simulates their scattering in underground detectors and the Earth overburden, is publicly available on the market.

\subsection{\micromegas: The FIMP regime}
The current \micromegas\, version v5.2 is a Fortran/C/C++ code which is fairly user-friendly: relic-density computations, direct detection and indirect detection predictions can all be switched on or off with simple flags in the main program. It is by far the most used tool in the dark matter community to study the phenomenology of WIMP generic models.

Besides the traditional $Z_2$ parity that is usually associated to dark matter particles (or $R$-parity in supersymmetry),  the relic density computation has been extended to cover the case of a dark group with a $Z_n$ symmetry; this implies to add into the Boltzmann equation terms accounting for semi-annihilation processes~\cite{Belanger:2014bga}, as well as the case of two DM candidates.

The most relevant novelty in \micromegas\, is the capability of computing the relic density of FIMPs via freeze-in. In short, the dark matter particles are too feebly interacting with SM particles in the early universe to be in chemical and kinetic equilibrium with the thermal bath. Dark matter is produced then either by the decay of an heavier (new) particles, either by the scattering of SM particles. Typically the initial dark matter number density is negligible, and it has been checked that the final value of the relic density is not very sensitive to this initial condition.  The approach followed by \micromegas\, tends to be as general as possible, namely based on only few assumptions, which should however be checked to be sure that they apply to the user model. \micromegas\,  considers three regimes for freeze-in: (i) dark matter production via the decay of a heavy mediator which is in thermal equilibrium, (ii) decay of a heavy mediator which is not in thermal equilibrium and (iii) $2 \to 2$ processes where SM particles produce two dark matter particles. It has been shown that the appropriate use of statistical distributions and thermal masses can lead to  differences up to factors of two, hence those have been properly included in \micromegas. In addition, the FIMP phenomenology is relevant for LHC searches, as it can give rise to long lived particles~\cite{Alimena:2019zri}, which can be considered using SModels~\cite{Ambrogi:2018ujg}.

\subsection{\maddm: Automatised annihilations for generic models}
The current \maddm\,version v3.1 is a mixed Python/Fortran code, which is based on the \mg\, platform and of which it inherits most features, such as the ability to compute any process automatically using the event generator \textsc{MadEvent}~\cite{Maltoni:2002qb}. This software has the capability to compute any annihilation cross-section into $n$ final state particles at tree level order. Additionally, \maddm\, is capable to compute loop-induced processes, by definition leading order hence finite loop processes, for the dark matter. This ability is based on \textsc{MadLoop}~\cite{Hirschi:2011pa}. A first attempt in the automatisation of loop-induced processes is presented in~\cite{Arina:2020udz}, however the code is currently being tested before its public release, which is foreseen for early 2021.  
The relevance of loop-induced processes is striking for dark matter observables: since the dark matter is neutral it can not couple directly to photons, hence the annihilation into di-photon can proceed only via loop (or effective vertex). Being a two body final state, each photon is monochromatic with an energy equal to the dark matter mass, since the annihilation occurs basically at rest. Such monochromatic lines, and in general sharp spectral features and edges in the gamma-ray energy spectra, are smoking gun for dark matter, as they can be hardly mimicked by astrophysical sources~\cite{Bouquet:1989sr,Bergstrom:1989jr,Rudaz:1989ij}. The other public codes can compute gamma-line predictions only for definite specific models, such as supersymmetry~\cite{Bergstrom:1989jr} or the inert doublet model~\cite{Gustafsson:2007pc}, as all diagrams contributing to the cross-section are hard-coded. \maddm\, will be able to compute on the fly such predictions for a generic dark matter model.

Still concerning the gamma-ray phenomenology, an interesting possibility to test models of dark sector featuring long-lived mediator particles is to observe the gamma-ray emission from the Sun due to dark matter annihilation, which has been proposed by several authors~\cite{Batell:2009zp,Leane:2017vag,Arina:2017sng} and has already evoked a search by the Fermi-LAT collaboration~\cite{Mazziotta:2020foa}. This module is not yet publicly available in \maddm\, but can be obtained by contacting the authors. 


\section{Discussion and conclusions}\label{sec:concl}
This review provides an (hopefully) unbiased overview of the status of the four major public tools for dark matter observables: \superiso, \darksusy, \micromegas\,  and \maddm. Concerning WIMPs, their phenomenology is very well covered in all of its necessary aspects; the user in this respect can simply choose the tool that better suits his/her needs and his/her model (files), knowing that the numerical discrepancies among tools are generically not larger than the percentage level. Concerning the new dark matter avenues, the dark sector phenomenology and sub-GeV dark matter particles begin to be pretty well described by \darksusy, while  \micromegas\, encompasses a fairly generic description of FIMPs. \maddm\, and \superiso\, will soon give the possibility to compute loop (induced) processes automatically within generic dark matter extensions of the SM. Last but not least, numerical tools are nowadays user friendly in both their usage and in their portability towards other tools for high energy physics. The use of \fr, \maddm, \micromegas, \mg, \ma, linked one after the other, has become a common practice in phenomenological papers, as well as the accomplishment of global fits is taking more and more momentum thanks to the modular tools developed by the GAMBIT collaboration.  Typically,  the confrontation of theoretical predictions with experimental data stands at the end of the chain for global analyses. A correct and sophisticated way of performing this comparison entails a likelihood analysis, as this confronts directly the model with the data. The exclusion bounds hence do not rely on approximations and assumptions that might not be verified for that particular model. The key ingredient is the experimental likelihood function: many efforts from the theoretical side is concentrated to reconstruct experimental likelihoods in order to recast properly limits from the collaborations. Usually, this is not a straightforward procedure when the data and the experimental details are not public. In this respect, in the past few years there has been a real improvement in the interactions between the experimental and theoretical communities and many data and likelihood functions have become public, such as for the Fermi-LAT dwarf spheroidal one. There are sectors where this is not yet common practice, but the efforts and synergies are moving to the good direction for data reinterpretation, see \eg~\cite{Abdallah:2020pec}.

Overall, this review has an optimistic view of the dark matter tool landscape. Nevertheless, this is not the end of the story and the public codes are continuously debugged, ameliorated and improved with new physics, as the quest for dark matter is not yet over. This calls for the continuous development of a platform for comprehensive studies and for the identification of new directions for the experimental searches to look for dark matter signatures.

\section*{Acknowledgments}
The author warmly acknowledges the organisers of TOOLS2020 for the invitation to give this review talk, which preparation has been enriching. The author is supported by the Innoviris  ATTRACT 2018 104 BECAP 2 agreement.

\bibliographystyle{JHEP}
\bibliography{biblio}

\end{document}